\documentclass[aps,nofootinbib,noshowpacs,showkeys,preprintnumbers,amsmath,amssymb]{revtex4}
\usepackage{graphicx,color}
\usepackage{graphics}
\usepackage{rotating}
\usepackage{amssymb}

\usepackage{color}

\usepackage{epsfig}
\usepackage{dcolumn}
\usepackage{bm}
\usepackage{multirow}

\def\be{\begin{equation}}
\def\ee{\end{equation}}
\def\ba{\begin{eqnarray}}
\def\ea{\end{eqnarray}}
\def\bea{\begin{eqnarray}}
\def\eea{\end{eqnarray}}
\def\bes{\begin{subequations}}
\def\ees{\end{subequations}}
\def\bear{\begin{array}}
\def\eear{\end{array}}

\newcommand{\A}{{\mathcal{A}}}
\newcommand{\tA}{{\widetilde {\mathcal{A}}}}
\newcommand{\ta}{{\widetilde a}}

\newcommand{\tk}{{\widetilde k}}

\newcommand{\MSbar}{\overline{\rm MS}}  
  
\newcommand{\LL}{\Lambda_{\rm L}}  

\newcommand{\tkap}{{\widetilde \kappa}}  

\begin{document}

\title{Towards unifying perturbative and Holographic Light-Front QCD via holomorphic coupling} 
\author{C\'esar Ayala$^a$}
\email{c.ayala86@gmail.com}
\author{Gorazd Cveti\v{c}$^b$}
\email{gorazd.cvetic@gmail.com}
\affiliation{$^a$Departamento de Ingenier\'ia y Tecnolog\'ias, Sede La Tirana, Universidad de Tarapac\'a, Av.~La Tirana 4802, Iquique, Chile}
\affiliation{$^b$Department of Physics, Universidad T{\'e}cnica Federico Santa Mar{\'\i}a, Avenida España 1680, Valpara{\'\i}so, Chile}

\date{\today}

\begin{abstract}
We construct a QCD coupling $\A(Q^2)$ in the Effective Charge (ECH) scheme of the canonical part $d(Q^2)$ of the (inelastic) polarised Bjorken Sum Rule (BSR) ${\overline \Gamma}_1^{{\rm p-n}}(Q^2)$. In the perturbative domain, the coupling $\A(Q^2)$ practically coincides with the perturbative coupling $a(Q^2)$ [$\equiv \alpha_s(Q^2)/\pi$] in the four-loop ECH renormalisation scheme. In the deep infrared (IR) regime, $\A(Q^2)$ behaves as suggested by the Holographic Light-Front QCD up to the second derivative. Furthermore, in contrast to its perturbative counterpart $a(Q^2)$, the coupling $\A(Q^2)$ is holomorphic in the entire complex $Q^2$-plane with the exception of the negative semiaxis, reflecting the holomorphic properties of the BSR observable $d(Q^2)$ [or: ${\overline \Gamma}_1^{{\rm p-n}}(Q^2)$] as dictated by the general principles of the Quantum Field Theory. It turns out that the obtained coupling, used as ECH, reproduces quite well the experimental data for ${\overline \Gamma}_1^{{\rm p-n}}(Q^2)$ in the entire $N_f=3$ regime $0 < Q^2 \lesssim 5 \ {\rm GeV}^2$.
\end{abstract}
\keywords{the strong coupling; deep inelastic scattering; resummation}
\maketitle

\section{Introduction}
\label{sec:intr}

The inelastic Bjorken polarised sum rule (BSR) \cite{BjorkenSR, BSR70} is the integral over the Bjorken-$x$ of the $g_1$ spin dependent structure functions of proton and neutron  
\be
{\overline \Gamma}_1^{{\rm p-n}}(Q^2)=\int_0^{1^{-}} dx \left[g_1^p(x,Q^2)-g_1^n(x,Q^2) \right].
\label{BSRdef}
\ee
Here, $Q^2 \equiv - q^2$ ($=-(q^0)^2 + {\vec q}^2$), where $q$ is the  is the momentum transfer given to the nucleon. In these experiments, $Q^2 > 0$, i.e., real spacelike regime. The upper integration limit $1^{-}$ in Eq.~(\ref{BSRdef}) excludes the $x=1$ singular (elastic) point.

The theoretical expression used for this quantity is usually Operator Product Expansion (OPE)
\be
\label{BSROPE}
{\overline \Gamma}_1^{{\rm p-n}, \rm {OPE}}(Q^2)= {\Big |}\frac{g_A}{g_V} {\Big |} \frac{1}{6} \left[ 1 - d(Q^2) - \delta d(Q^2)_{m_c} \right] +\sum_{i=2}^\infty \frac{\mu_{2i}(Q^2)}{Q^{2i-2}}.
\ee
In the leading-twist (dimension $D=0$) part, $|g_A/g_V|$ is the ratio of the nucleon axial charge, and we take the value $|g_A/g_V|=1.2754$ \cite{PDG2023}.
Further, $d(Q^2)$ is the canonical QCD-part of BSR in the leading-twist ($D=0$) OPE contribution, and this $d(Q^2)$ is regarded here formally as the $N_f=3$ perturbative QCD (pQCD) contribution, i.e., the perturbative contribution of the massless QCD with three flavours ($u$, $d$ and $s$). The quantity $\delta d(Q^2)_{m_c}$ is the small correction due to the nondecoupling effects of the $c$-quark, i.e., the effects from $m_c \not= \infty$; it is known only at the leading order \cite{Blumetal}, $\delta d(Q^2)_{m_c} \sim a^2$. 

There are many experimental results for ${\overline \Gamma}_1^{{\rm p-n}}(Q^2)$, especially in the region of low $Q^2$ ($Q^2 < 2 \ {\rm GeV}^2$). They are from different experiments: CERN \cite{CERN}, DESY \cite{DESY}, SLAC \cite{SLAC}, and Jefferson Lab \cite{JeffL1,JeffL2,JeffL3,JeffL4,JeffL5}. These results also include a few high values $Q^2 > 4.74 \ {\rm GeV}^2$, which we will not include in our analysis because the construction of our theoretical expression will be for $N_f=3$ active massles quark flavours.\footnote{
The extension of our construction to $N_f=4$ is challenging and we will not consider this case here.}

Many theoretical evaluations of the BSR have been performed in the literature. An incomplete list includes the works \cite{ABFR,JeffL1,JeffL2,JeffL4,ACKS,KoMikh,LFHBSR,BSRPMC,AyPi} where pQCD coupling was used, $a(\mu^2) \equiv \alpha_s(\mu^2)/\pi$, in the canonical part $d(Q^2)$ of BSR. In \cite{pPLB} we used a renormalon-motivated resummation of BSR with pQCD coupling. We point out that pQCD running couplings $a(\mu^2)=a(\kappa Q^2)$ have mathematical artifacts called Landau singularities for positive small $\mu^2$ (or: $Q^2$), which limit the applicability to the evaluation of  $d(Q^2)$ to high $Q^2$. In \cite{ACKS} we used for BSR $d(Q^2)$ truncated series in terms of holomorphic variants (AQCD) of the coupling, $a \mapsto \A$, i.e., couplings that have no Landau singularities in the complex $Q^2$-plane (there are only singularities on the $Q^2<0$ timelike semiaxis). Specifically, we used the 2$\delta$AQCD \cite{2dAQCD,2dAQCDb} and 3$\delta$AQCD couplings \cite{3dAQCD,3dAQCDb} in the evaluation of $d(Q^2)$ in \cite{ACKS}. On the other hand, in \cite{pNPB,pNPPP} we used such couplings in the renormalon-motivated resummation approach. In the works \cite{Gabd} the authors applied the Minimal Analytic (MA) version \cite{Shirkov,SMS} of holomorphic coupling to the truncated series of BSR $d(Q^2)$.

There is another approach to the evaluation of BSR, namely use of the Holographic Light-Front QCD (HLFQCD) effective coupling \cite{LFH1}. This coupling $\A(Q^2)$ has a known behaviour for $Q^2 \to 0$, namely of the form $\exp(-Q^2/\tkap^2)$ (where $\tkap \approx 0.523$ GeV); however, in other $Q^2$-regimes, this approach does not predict the coupling. Therefore, a priori, it is not clear how to extend this coupling to the entire $Q^2$-plane in such a way that the coupling would not have any Landau singularities (and would thus reflect the holomorphic properties of spacelike QCD observables) and would simultaneously behave as a pQCD coupling for large $|Q^2| > 1 \ {\rm GeV}^2$. In Refs.~\cite{LFHm1,LFHm2,LFHm3} a matching procedure was constructed between the low-$Q^2$ Gaussian behaviour and the high-$Q^2$ perturbative behaviour (in various schemes), by matching at an intermediate positive $Q_0^2$ ($\sim 1 \ {\rm GeV}^2$). The matching was performed for the values of the coupling and its (first) derivative. The perturbative coupling was taken to run with four-loop $\beta$-function (i.e., including the terms $\propto \beta_3$). The resulting coupling $\A(Q^2)$ and its (first) derivative were then continuous on the positive axis. However, the coupling was not holomorphic, because of the discontinuity of higher derivatives of $\A(Q^2)$ at the spacelike $Q^2=Q_0^2$, i.e., (Landau) singularity there.

This problem of the mentioned discontinuity at positive $Q^2=Q_0^2$ was addressed in the work \cite{LFHext}, where the extension to high positive $Q^2$ was made by avoiding singularity of higher derivatives at a positive $Q_0^2$. The resulting perturbative coupling in the high-$Q^2$ regime runs according to the one-loop RGE formula. Nonetheless, two complex-conjugate singularities appear in the $Q^2$-plane outside the real axis, and they are again Landau singularities in the generalised sense, because the $Q^2$-complex plane without the timelike axis, $\mathbb{C} \backslash (-\infty, 0]$, is regarded as spacelike, and all spacelike QCD observables ${\cal D}(Q^2)$ (such as BRS, Adler function, etc.) are holomorphic functions of $Q^2$ in the entire spacelike $Q^2$-complex plane, i.e., for all $Q^2 \in \mathbb{C} \backslash (-\infty, 0]$. For these reasons, it is desirable to construct a running coupling $\A(Q^2)$ with analogous holomorphic properties, i.e., without any Landau singularities. 

  In our work we address this problem from a somewhat different point of view. We find for the canonical part of BSR, $d(Q^2)$, the four-loop pQCD ECH scheme, i.e., such a scheme in which the coupling $a(\kappa_{\rm ECH} Q^2; \beta_2^{\rm ECH}, \beta_3^{\rm ECH})$ coincides with $d(Q^2)$ up to (and including) $\sim a^4$. This (pQCD) coupling has Landau singularities. Then we construct a holomorphic coupling $\A(Q^2)$  [$a(Q^2) \mapsto \A(Q^2)$], i.e., without Landau singularities, which agrees with $a(Q^2; \beta_2^{\rm ECH}, \beta_3^{\rm ECH})$ for large $Q^2$, and at low $Q^2 \to 0$ its value and the first two derivatives agree with those of the HLCQCD coupling $\propto \exp(-Q^2/\tk^2)$. We point out that the coupling constructed in the present work is different from the 2$\delta$QCD \cite{2dAQCD,2dAQCDb} and 3$\delta$AQCD \cite{3dAQCD,3dAQCDb} couplings, because the latter couplings, by its form, cannot fulfill all the mentioned low-$Q^2$ conditions (while they do fulfill the high-$Q^2$ conditions).

In Sec.~\ref{sec:constr} we construct such a holomorphic coupling. We refer for details to Appendices \ref{app:RSch}, \ref{app:P44} and \ref{app:Acond}, and for the (small) charm-quark nondecoupling corrections to Appendix \ref{app:ddmc}. In Sec.~\ref{sec:res} we present the numerical results of the obtained ECH holomorphic approach and compare it with the experimental BSR data. In Sec.~\ref{sec:summ} we summarise our results.

\section{Construction}
\label{sec:constr}

In the nonperturbative sense, we will regard the ($N_f=3$) effective charge (ECH) of the canonical part $d(Q^2)$ in BSR Eq.~(\ref{BSROPE}) as such a quantity $d(Q^2)_{\rm ECH}$ that appears in the OPE Eq.~(\ref{BSROPE}) by replacing $d(Q^2)$ and, at the same time, is regarded as containing all the nonperturbative ($D=2,4,\ldots$) OPE contributions, for all $Q^2$ (including $Q^2 \to 0$):
\be
\label{BSRECH}
{\overline \Gamma}_1^{{\rm p-n}, \rm {ECH}}(Q^2)= {\Big |}\frac{g_A}{g_V} {\Big |} \frac{1}{6} \left[ 1 - d(Q^2)_{\rm ECH} - \delta d(Q^2)_{m_c} \right].
\ee

First we summarise the ECH scheme in pQCD, following the ideas of Grunberg and others \cite{ECH1,ECH2,ECH3}.

The perturbation expansion of $d(Q^2)$ in powers of the QCD coupling $a(\kappa Q^2) \equiv \alpha_s(\kappa Q^2)/\pi$ is at present known exactly up to order $a^4$ \cite{GorLar1986,LarVer1991,BaiCheKu2010}
\bea
d(Q^2)_{\rm pt} & = & a(\kappa Q^2) + d_1(\kappa) a(\kappa Q^2)^2 + d_2(\kappa; c_2)  a(\kappa Q^2)^3 + d_3(\kappa; c_2, c_3) a(\kappa Q^2)^4 + {\cal O}(a^5),
\label{dptkap}
\eea
where $\kappa=\mu^2/Q^2$ is the renormalisation scale (RScl) parameter ($\mu^2$ is squared RScl). In the above (exactly known) coefficients $d_j$ we have the dependence on the scheme via the RScl parameter $\kappa$ and the scheme parameters $c_j \equiv \beta_j/\beta_0$ ($j=2,3$), where these scheme parameters appear in the $\beta$-function
\be
\frac{d}{d \ln \mu^2} a(\mu^2) = - \beta_0 a(\mu^2)^2 \left[ 1 + \sum_{j \geq 1} c_j  a(\mu^2)^{j} \right]
\label{RGE} \ee
The first two coefficients, $\beta_0=(11- (2/3) N_f)/4$ and $c_1 =(102 - (38/3) N_f)/16/\beta_0$, are universal in the mass independent schemes.
The coupling $a(\kappa Q^2)$ depends on $\kappa$ and all the scheme parameters $c_j$ ($j \geq 2$). The dependence of the coefficients $d_j$ on $\kappa$ and $c_j$'s is then fixed by requiring that $d(Q^2)$ is independent of those parameters. These dependencies are compiled in Appendix \ref{app:RSch}. The resulting numerical values of the ECH pQCD scheme (with $N_f=3$) from there, according to Eqs.~(\ref{ECHval}), are
\be
\kappa_{\rm ECH}=0.203398; \quad c_2^{\rm ECH}=5.47568; \quad c_3^{\rm ECH}=112.690.
\label{RSchECH} \ee
This can be compared with the canonical ($N_f=3$) $\MSbar$ scheme values: $\kappa=1$, ${\bar c}_2=4.47106$ and ${\bar c}_3=20.9902$.
Stated otherwise, we have in pQCD:
\be
d(Q^2) = a(\kappa_{\rm ECH} Q^2; c_2^{\rm ECH},  c_3^{\rm ECH}) + {\cal O}(a^5),
\label{dECHpQCD} \ee
i.e., the rescaling $Q^2 \mapsto \kappa_{\rm ECH} Q^2$ and the change of scheme parameters ${\bar c}_j \mapsto c_j^{\rm ECH}$ ($j=2,3$) in the pQCD coupling absorbs the power terms in Eq.~(\ref{dptkap}) up to (and including) $\sim a^4$.

As expected, the pQCD coupling $a(\kappa_{\rm ECH} Q^2; c_2^{\rm ECH},  c_3^{\rm ECH})$ has Landau singularities at low positive $Q^2$. For example, when $\alpha_s^{\MSbar}(M_Z^2)=0.1179$, we have $a(\kappa_{\rm ECH} Q^2; c_2^{\rm ECH},  c_3^{\rm ECH})$ complex (nonreal) for $Q^2 \leq 5.262 \ {\rm GeV}^2$  ($\sqrt{Q^2} < 2.294$ GeV), i.e., there is a large Landau cut $0 \leq Q^2 \leq 5.262 \ {\rm GeV}^2$ in the complex $Q^2$-plane.

Now we proceed in the following way. In the above scheme, we construct a corresponding holomorphic (ECH) coupling, $a(\kappa_{\rm ECH} Q^2; c_2^{\rm ECH},  c_3^{\rm ECH}) \mapsto \A(\kappa_{\rm ECH} Q^2; c_2^{\rm ECH},  c_3^{\rm ECH})$. Stated otherwise, $\A(Q^2; c_2^{\rm ECH},  c_3^{\rm ECH}) \equiv \A(Q^2)$ is a coupling that has no Landau singularities, i.e., it is a holomorphic function in the $Q^2$-complex plane with the exception of the timelike semiaxis $Q^2 < 0$, reflecting thus the analytic (holomorphic) properties of QCD observables such as ${\overline \Gamma}_1^{\rm p-n}(Q^2)$ [$\Leftrightarrow \; d(Q^2)$] as a function in the complex $Q^2$-plane. Further, we require that this coupling practically coincide with the underlying pQCD coupling $a(\kappa_{\rm ECH} Q^2; c_2^{\rm ECH},  c_3^{\rm ECH}) \equiv a(Q^2)$ for sufficiently large $|Q^2| > 1 \ {\rm GeV}^2$. In practice, we require
\be
\A(Q^2) - a(Q^2) \sim (\LL^2/Q^2)^5  \quad (|Q^2| > 1 \ {\rm GeV}^2).
  \label{diffAa} \ee
We will see that the high-momentum condition (\ref{diffAa}) represents four conditions for the parameters of the low-energy regime of the spectral function $\rho_{\A}(\sigma) = {\rm Im} \; \A(-\sigma - i \epsilon)$. [We note that $\rho_{\A}(\sigma)$ appears in the dispersion representation of the coupling $\A(Q^2)$, see Eqs.~(\ref{Af}).]

 At low $|Q^2| < 1 \ {\rm GeV}^2$, i.e., when $Q^2 \to 0$, we require that the coupling $\A(Q^2)$ behave as suggested by the Holographic Light-Front (HLF) QCD \cite{PLB759}:
\be
\A(\kappa_{\rm ECH} Q^2) \approx \exp \left( - \frac{Q^2}{4 \tkap^2} \right),
\qquad (|Q^2| < \tkap^2),
\label{AQto0} \ee
where $\tkap \approx 0.523$ GeV \cite{PLB759}. It turns out that in our approach it is in practice very difficult (or perhaps impossible) to implement the low-momentum LFH-condition (\ref{AQto0}) exactly. Therefore, we will implement it in the following approximation
\be
\A(\kappa_{\rm ECH} Q^2)  \approx 1 - \frac{Q^2}{4 \tkap^2} + \frac{1}{2!} \left( \frac{Q^2}{4 \tkap^2} \right)^2 +  {\cal O}\left( \left( \frac{Q^2}{\tkap^2} \right)^3 \right),
\label{AQto0app} \ee
i.e., we will implement three low-momentum conditions
\bes
\label{AQ0}
\bea
\A(0) = 1, \quad \frac{d}{d Q^2}\A(Q^2){\bigg |}_{Q^2=0} &=&- \frac{1}{4 \tkap^2 \kappa_{\rm ECH}},
\label{AdAQ0}
\\
\left(\frac{d}{d Q^2}\right)^2 \A(Q^2){\bigg |}_{Q^2=0} &=& \left( \frac{1}{4 \tkap^2 \kappa_{\rm ECH}} \right)^2.
\label{ddAQ0}
\eea \ees
Thus, altogether we have seven conditions: four from the high-momentum regime, and three from the low-momentum regime.

If we introduce the dimensionless momenta scaled by the Landau\footnote{The value of the Landau scale $\LL$ ($\sim 0.1$ GeV) of the $N_f=3$ regime is determined by the value of $\alpha_s^{\MSbar}(M_Z^2)$ (which is from the $N_f=5$ regime), as explained in Appendix \ref{app:P44}.} scale $\LL$, namely $u \equiv Q^2/\LL^2$ and $s= \sigma/\LL^2$, our ansatz for the spectral function $r_{\rm A}(s) \equiv \rho_{\A}(\sigma) = {\rm Im} \; \A(-\sigma - i \epsilon)$ is
\bea
\frac{1}{\pi} r_{\A}(s) & = & f_1^{(0)} \delta(s - s_1) + f_1^{(1)} \delta^{'}(s-s_1) + f_1^{(2)} \delta^{''}(s-s_1) + f_1^{(3)} \delta^{'''}(s-s_1)
\nonumber\\ &&
+ f_2^{(0)} \delta(s - s_2) + \frac{1}{\pi} \Theta(s - s_0) r_{a}(s),
\label{rA} \eea
where $r_{a}(s) = \rho_a(\sigma) = {\rm Im} \; a(-\sigma - i \epsilon)$ is the spectral function of the underlying pQCD coupling. It is assumed that $0 < s_1 < s_2< s_0$, so that the discontinuity function $r_{\A}(s)$ is zero for $s < 0$ ($s<s_1$), i.e., there are no Landau singularities of $\A(Q^2)$ on the negative $Q^2$-semiaxis. The idea of the ansatz (\ref{rA}) is similar to that of the 2$\delta$AQCD \cite{2dAQCD,2dAQCDb} and 3$\delta$AQCD \cite{3dAQCD,3dAQCDb} holomorphic couplings. The spectral function  $r_{\A}(s)$ coincides with the corresponding pQCD spectral function $r_a(s)$ at high scales $s > s_0$. At lower scales $s< s_0$ the (otherwise unknown) behaviour of $r_{\A}(s)$ is parametrised by a combination of two Dirac deltas (at $s=s_1$ and $s=s_2$) and three derivatives of delta, all at the same low scale $s=s_1$. Altogether, we have eight real parameters: $f_1^{(k)}$ ($k=0,\ldots,3$), $f_2^{(0)}$, $s_j$ ($j=1,2,0$). They are fixed by the seven aforementioned conditions, and by a judicial choice of the values of the eighth parameter. The seven conditions are written down explicitly in Appendix \ref{app:Acond}.

More specifically, the holomorphic running coupling $f(u) \equiv \A(Q^2)$ is obtained from the spectral function $r_{\A}(s) \equiv \rho_{\A}(\sigma)$ by the usual dispersion relation based on Cauchy theorem
\bes
\label{Af}
\bea
\A(Q^2) &=& \frac{1}{\pi} \int_0^{\infty} d \sigma \frac{\rho_{\A}(\sigma)}{(\sigma + Q^2)}, \; \Rightarrow
\label{Adisp}
\\
f(u) \equiv \A(u \LL^2) & = & \frac{1}{\pi} \int_0^{\infty} ds \frac{r_{\A}(s)}{(s + u)}
\label{fdisp}
\\
& = & \sum_{k=0}^3 \frac{f_1^{(k)} k!}{(s_1+u)^{k+1}} + \frac{f_2^{(0)}}{(s_2+u)}
  + \frac{1}{\pi} \int_{s_0}^{+\infty} ds \frac{r_{a}(s)}{(s + u)}.
\label{Aexpr} \eea \ees
  
We took in our numerical analysis the value $\alpha_s^{\MSbar}(M_Z^2)=0.1179$, the central world average value \cite{PDG2023}. This value fixes the Landau scale $\LL$ as explained in Appendix \ref{app:P44}. As explained in Appendix \ref{app:Acond}, we have seven conditions imposed on the coupling, which fix seven out of the eight aforementioned (dimensionless) parameters of the coupling. This leaves us with one free parameter, which we choose to be $s_1$, or equivalently, the threshold momentum scale $M_1 = \sqrt{s_1} \LL$. On physical grounds, we expect that this threshold scale is comparable to the smallest hadronic (threshold) scales. Therefore, we choose this scale to be $M_1 = (2 m_{\pi})$ ($\approx 0.279$ GeV), and we vary this scale in the interval $(2 m_{\pi})/\sqrt{2} < M_1 < (2 m_{\pi}) \sqrt{2}$. It turns out that we do get solutions to the seven conditions when the chosen value of $M_1$ is increased all the way up to $M_1= (2 m_{\pi}) \sqrt{2}$ ($\approx 0.395$ GeV). On the other hand, when decreasing the threshold $M_1$, the smallest value of $M_1$ which still gives solutions for the seven parameters is $M_1 \approx 0.214$ GeV. Therefore, our variation of the threshold scale is $M_1 = (0.279^{+0.116}_{-0.065})$ GeV (we note that $M_1=\sqrt{s_1} \LL$). The resulting values of the parameters $s_j$, $f_1^{(k)}$ and $f_2^{(0)}$ are given in Appendix \ref{app:Acond} in Table \ref{TabParam}. 

Having fixed the parameters of the coupling $\A(Q^2)$, we thus obtain
\be
d(Q^2)_{\rm ECH} = \A(\kappa_{\rm ECH} Q^2).
\label{dECH} \ee
The (small) correction $\delta d(Q^2)_{m_c}$, appearing in Eqs.~(\ref{BSROPE})-(\ref{BSRECH}), is evaluated as explained in Appendix \ref{app:ddmc}. There, we also see that $\delta d(Q^2)_{m_c}$ at small $Q^2$ ($< 0.3 \ {\rm GeV}^2$) goes to zero fast when $Q^2 \to 0$, cf.~Figs.~\ref{Figddmc}. The use of the quantity Eq.~(\ref{dECH}) and the (small) correction $\delta d(Q^2)_{m_c}$ Eq.~(\ref{deldA}) in the formula (\ref{BSRECH}) then gives us our prediction of the (inelastic) BSR ${\overline \Gamma}_1^{{\rm p-n}}(Q^2)$.

\section{Results}
\label{sec:res}

\begin{figure}[htb] 
\begin{minipage}[b]{.49\linewidth}
\includegraphics[width=80mm,height=50mm]{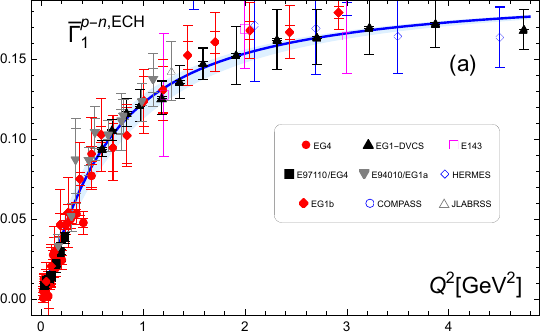}
\end{minipage}
\begin{minipage}[b]{.49\linewidth}
  \includegraphics[width=80mm,height=50mm]{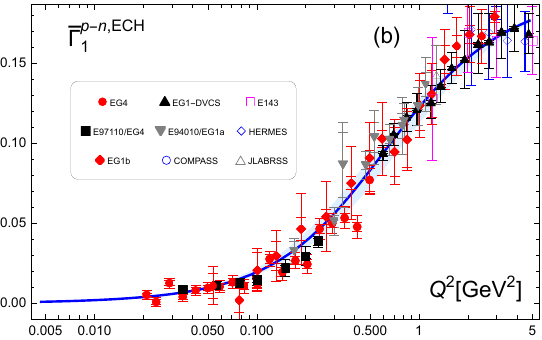}
\end{minipage} 
 \caption{\footnotesize (a) The ECH values of BSR ${\overline \Gamma}_1^{p-n}(Q^2)$ in the holomorphic ECH approach, and the corresponding experimental data. The threshold scale is taken to be $M_1 = (0.279^{+0.116}_{-0.065})$ GeV, where the central (solid) line is for $M_1 = 0.279$ GeV, and the upper and the lower borders of the grey stripe are for $M_1=0.214$ GeV and $M_1=0.345$ GeV, respectively. (b) The same as Fig.~(a), but with $Q^2$ scaled logarithmically. \textcolor{black}{The experimental data are included.}}
\label{FigBSR}
\end{figure}
\begin{figure}[htb] 
\begin{minipage}[b]{.49\linewidth}
\includegraphics[width=80mm,height=50mm]{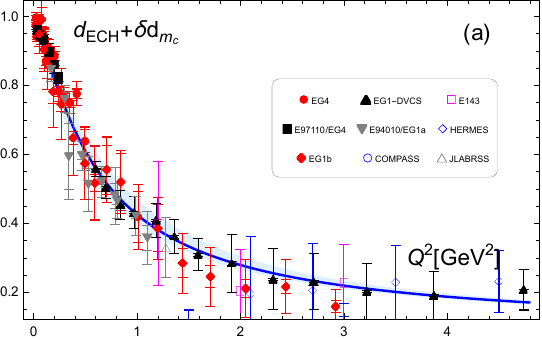}
\end{minipage}
\begin{minipage}[b]{.49\linewidth}
  \includegraphics[width=80mm,height=50mm]{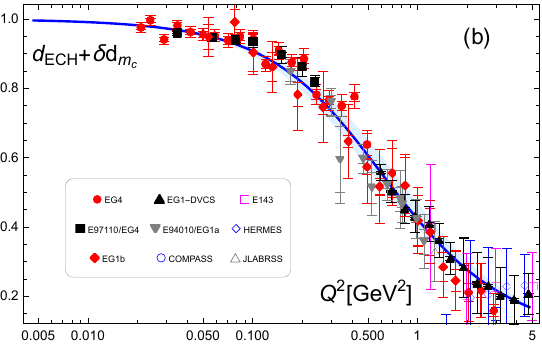}
\end{minipage} 
 \caption{\footnotesize The same as in the previous Figure \ref{FigBSR}, but now for the canonical part $d(Q^2)_{\rm ECH} +\delta d(Q^2)_{m_c}$.}
\label{FigBSRc}
\end{figure}
In Fig.~\ref{FigBSR} we present the resulting BSR ${\overline \Gamma}_1^{p-n}(Q^2)$ evaluated with our holomorphic ECH approach Eq.~(\ref{BSRECH}), together with the experimental data. In Fig.~\ref{FigBSRc} we present, alternatively, the canonical quantity $d(Q^2)_{\rm ECH} + \delta d(Q^2)_{m_c}$, together with the corresponding experimental data. These Figures are for the interval $0 \leq Q^2 < 4.74 \ {\rm GeV}^2$, i.e., for the regime where we can apply $N_f=3$ massless QCD ($d(Q^2)_{\rm ECH}$) and small $\delta d(Q^2)_{m_c}$ correction. Comparison of Fig.~\ref{FigBSRc} (here) with Fig.~\ref{Figddmc}(b) (in Appendix \ref{app:ddmc}) shows that the charm quark nondecoupling ($m_c \not= \infty$) correction $\delta d(Q^2)_{m_c}$ is several orders of magnitude smaller than $d(Q^2)_{\rm ECH}$ in the entire considered $Q^2$-interval. 

\begin{figure}[htb] 
\begin{minipage}[b]{.49\linewidth}
\includegraphics[width=80mm,height=50mm]{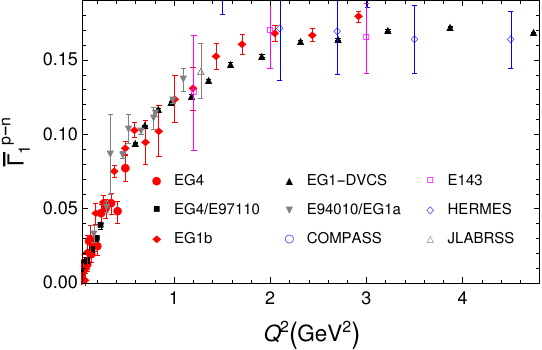}
\end{minipage}
\begin{minipage}[b]{.49\linewidth}
  \includegraphics[width=80mm,height=50mm]{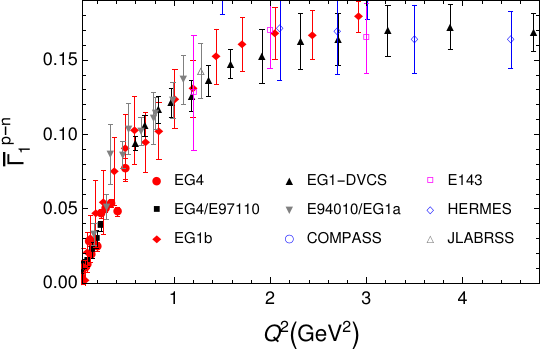}
\end{minipage} 
\caption{\footnotesize The measured values for the inelastic BSR ${\overline \Gamma}_1^{p-n}(Q^2)$ for various experiments: the left Figure is with the statistical and the right Figure is with the systematic uncertainties as vertical bars. The Figures were taken from \cite{pPLB}.}
\label{Figstasys}
\end{figure}
In Figs.~\ref{FigBSR} and \ref{FigBSRc} we included also the experimental results for (inelastic) BSR as measured by various experiments: CERN \cite{CERN}, DESY \cite{DESY}, SLAC \cite{SLAC}, and Jefferson Lab \cite{JeffL1,JeffL2,JeffL3,JeffL4,JeffL5}. The statistical and systematic experimental uncertainties are superimposed there (with overlapping). For convenience, we present in Figs.~\ref{Figstasys} our compilation \cite{pPLB} of these measured values, separately with statistical and systematic uncertainties. They are organised into various subsets from various detectors: (a) E94010 (Jefferson Lab (JL) Hall A); (b) EG1a, EG1b, EG1-DVCS, EG4 (JL Hall B); JLABRSS (JL Hall C); HERMES (DESY); E143 (SLAC); COMPASS (CERN). \textcolor{black}{The compilation of these data, for the considered $Q^2$-interval $0 < Q^2 < 4.74 \ {\rm GeV}^2$, is also available on www \cite{BSRexp}.}

We can see in Figs.~\ref{FigBSR} and \ref{FigBSRc} that the quality of our ECH solution, at least for the central choice of threshold scale $M_1=2 m_{\pi}$ ($=0.279$ GeV), is good for the entire considered $Q^2$-interval $0 \leq Q^2 \leq 4.793 \ {\rm GeV}^2$. We can quantify this, by considering the $\chi^2$-quantity
\be
\chi^2(j_{\rm min}; k) = \frac{1}{(j_{\rm max}-j_{\rm min}+1)}\sum_{j=j_{\rm min}}^{j_{\rm max}} \frac{ \left[ {\overline \Gamma}_1^{{\rm p-n},{\rm ECH}}(Q_j^2) - {\overline \Gamma}_1^{{\rm p-n}}(Q_j^2)_{\rm exp} \right]^2}{\sigma(Q_j^2; k)^2}.
\label{chi2} \ee
Here, we have 77 points of experimental data, they are the discrete scales $Q_j^2$: $Q_1^2 < \cdots < Q_{77}^2$, where $Q_1^2=0.021 \ {\rm GeV}^2$ and $Q_{77}^2=4.739 \ {\rm GeV}^2$. Thus, in Eq.~(\ref{chi2}), $j_{\rm max}=77$, and $j_{\rm min}=1$. In the expression (\ref{chi2}) we have the uncorrelated squared uncertainties $\sigma(Q_j^2; k)^2$ at $Q_j^2$, and they are in principle unknown. While the statistical errors $\sigma_{\rm stat}(Q_j^2)$ are uncorrelated, the systematic errors $\sigma_{\rm sys}(Q_j^2)$ may have some (unknown) correlations. In \cite{pPLB,pNPB} we argued, using the method of unbiased estimate \cite{Deuretal2022,PDG2020,Schmell1995}, that these combined uncorrelated (squared) uncertaintites $\sigma(Q_j^2; k)^2$ are
\be
\sigma^2(Q_j^2; k) = \sigma^2_{\rm stat}(Q_j^2) + k \; \sigma^2_{\rm sys}(Q_j^2),
\label{sig2} \ee
with $k \approx 0.15$. If we take $k=0.15$, we obtain for our ECH solution the values
\be
\chi^2(j_{\rm min}=1; 0.15) = 1.676^{+1.252}_{+1.119} \quad
\left( M_1 = \left( 0.279^{+0.116}_{-0.065} \right) \; {\rm GeV} \right).
\label{xi2num} \ee
\textcolor{black}{We note that both signs in the variation of $\chi^2$ above are positive, indicating that our central choice $M_1=0.279$ GeV gives approximately the minimal value of $\chi^2$ under the variation of $M_1$.}
In \cite{pNPB} we applied AQCD variants (2$\delta$AQCD and 3$\delta$AQCD) with truncated OPE expansion of BSR (and a renormalon-motivated resummation of $d(Q^2)$) and performed fits in the interval $0.592 \ {\rm GeV}^2 \leq Q^2 \leq 4.739 \ {\rm GeV}^2$, which corresponds to $j_{\rm min}=40$ (and $j_{\rm max}=77$).\footnote{
The renormalisation schemes for the applied 2$\delta$AQCD \cite{2dAQCD,2dAQCDb} and 3$\delta$AQCD \cite{3dAQCD,3dAQCDb} variants there were of the type of P44 with given $c_2$ and $c_3$ values, we refer for details to \cite{pNPB}.}
In this shorter $Q^2$-interval ($j_{\rm min}=40$), and setting $k=0.15$, the expression (\ref{chi2}) gives us\footnote{
We point out that in \cite{pNPB}, the OPE (\ref{BSROPE}) was taken truncated at either the $D=2$ term $\sim 1/Q^2$ (one-parameter fit, $N_{\rm p}=1$) or the $D=4$ term $\sim 1/(Q^2)^2$ (two-parameter fit, $N_{\rm p}=2$). The denominator in $\chi^2$ Eq.~(\ref{chi2}) in front of the sum was $(j_{\rm max}-j_{\rm min}+1 - N_{\rm p})$ (with $j_{\rm max}=77$ and $j_{\rm min}=40$), and the $k$ factor was adjusted so that $\chi^2=1$. If we change the denominator to that in the expression (\ref{chi2}) here and set $k=0.15$ (but still keep $j_{\rm max}=77$ and $j_{\rm min}=40$), the results of the fit procedure of Ref.~\cite{pNPB} then give us for such $\chi^2$ the values $\chi^2=0.898$ and $0.890$ for 2$\delta$AQCD (and one- or two-parameter fit, respectively), and $\chi^2=1.004$  for 3$\delta$AQCD (and one- or two-parameter fit).}
\be
\chi^2(j_{\rm min}=40; 0.15) = 0.938^{+3.025}_{+0.261} \quad
\left( M_1 = \left( 0.279^{+0.116}_{-0.065} \right) \; GeV \right).
\label{xi2num2} \ee
The results (\ref{xi2num}) and (\ref{xi2num2}) thus suggest that we have a high quality coincidence of the considered theoretical ECH result with the experimental data, at least for the central choice $M_1=2 m_{\pi}$ ($=0.279$ GeV).

\begin{figure}[htb] 
\begin{minipage}[b]{.49\linewidth}
\includegraphics[width=80mm,height=50mm]{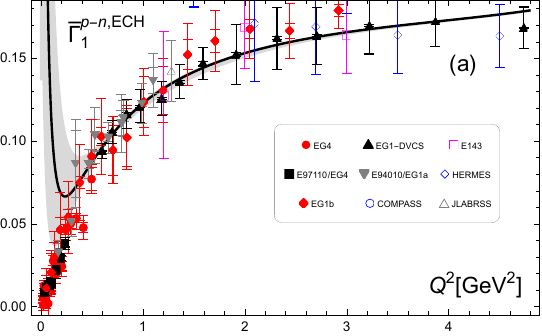}
\end{minipage}
\begin{minipage}[b]{.49\linewidth}
  \includegraphics[width=80mm,height=50mm]{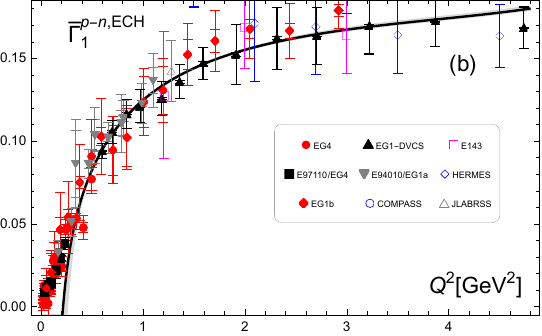}
\end{minipage} 
\caption{\footnotesize \textcolor{black}{(a) As Fig.~\ref{FigBSR}(a), but for the one-parameter fit of the OPE (\ref{BSROPE}), using in the leading-twist the ECH coupling Eqs.~(\ref{BSRECH}), (\ref{dECH}); the solid line corresponds to the central value of ${\bar f}_2$ Eq.~(\ref{bf21p}), and the grey band corresponds to the variation of ${\bar f}_2$ Eq.~(\ref{bf21p}); (b) The same as in (a), but for the two-parameter fit, with the parameter values Eqs.~(\ref{bf2mu6}).}}
\label{FigBSROPE}
\end{figure}

\begin{figure}[htb] 
\centering\includegraphics[width=80mm]{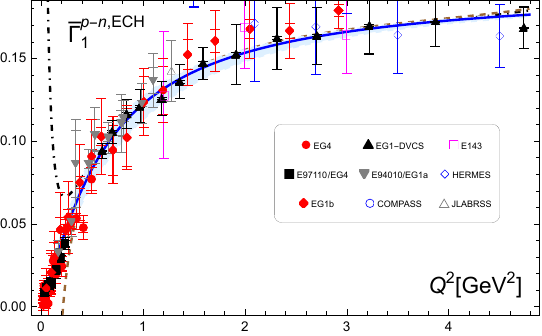}
 \vspace{4pt}
 \caption{\footnotesize \textcolor{black}{The central OPE curves of Fig.~\ref{FigBSROPE} together with the curve and stripe of the pure ECH approach [the curve and stripe of Fig.~\ref{FigBSR}(a)]. The one-parameter (two-parameter) fit OPE curve is dash-dotted (dashed).}}
\label{FigBSRECHvsOPE}
\end{figure}

\begingroup\color{black}
For comparisons with the above (pure) ECH results and with the OPE-fit results of Ref.~\cite{pNPB}, it may be interesting to apply the considered ECH expression Eq.~(\ref{BSRECH}) combined with the additional $D$( $\equiv 2 i - 2$)$=2,4$ terms of the OPE expansion Eq.~(\ref{BSROPE}), and perform a completely analogous fit to the data as we performed it in \cite{pNPB}.\footnote{\textcolor{black}{We point out that, in contrast, in \cite{pNPB} the 2$\delta$AQCD and 3$\delta$AQCD holomorphic couplings were used, and the canonical part $d(Q^2)$ was evaluated by a renormalon-motivated resummation, while here $d(Q^2)$ is simply the considered ECH coupling Eq.~(\ref{dECH}).}}
The fit is performed in a restricted interval $Q^2_{\rm min} \leq Q^2 \leq 4.74 \ {\rm GeV}^2$. The value of $Q^2_{\rm min}$, and of the $k$-parameter of $\sigma^2$ [cf.~Eq.~(\ref{sig2})] that appears in $\chi^2$, are determined by a procedure described in \cite{pNPB}. For the central coupling case of $M_1=0.279$ GeV we obtain $Q^2_{\rm min} = 0.592 \ {\rm GeV}^2$ (this happens to be the same value as in \cite{pNPB}), and the value of $k$ is $k=0.1190$ for the two-parameter fit (i.e., with the $D=2$ and $D=4$ terms) and $k=0.1353$ for the one-parameter fit (i.e., when only the $D=2$ term is included). These values of $k$ are somewhat different from those obtained in \cite{pNPB}. When we vary $M_1$ around the mentioned central value as before, i.e., $M_1 = \left( 0.279^{+0.116}_{-0.065} \right)$ GeV, but keep the mentioned $Q^2_{\rm min}$ and $k$ values fixed for simplicity, we obtain the following values of the extracted fit parameters:
\bes
\label{bf2mu6}
\bea
{\bar f}_2 & = & +0.0008^{+0.0887}_{-0.0312},
\label{bf2}
\\
\mu_6 & = & -0.0056^{-0.0074}_{+0.0015} \; [{\rm GeV}^4],
\label{mu6} \eea \ees
and for the one-parameter fit
\bea
\label{bf21p}
{\bar f}_2 & = & -0.0408^{+0.0343}_{-0.0203}.
\eea      
The central values of these parameters are small, indicating that the considered ECH coupling approach does not need significant corrections from the $D=2,4$ terms. The values of these parameters are roughly of the same order of magnitude as those obtained in \cite{pNPB} from the fit using the 2$\delta$AQCD coupling, and are significantly smaller than those obtained in \cite{pNPB} by the fit using the 3$\delta$AQCD coupling. We recall that in all the OPE-fit cases, the resulting $\chi^2$ attains the value $\chi^2=1.$. However, we point out that in contrast to the pure ECH approach presented in this work, the OPE approach must fail at low $Q^2$, and that is why we have the exclusion of a significant portion of data (namely, for $0 < Q^2 < 0.592 \ {\rm GeV}^2$) from the OPE fits.

In Figs.~\ref{FigBSROPE} we present, in analogy with the pure ECH case of Fig.~\ref{FigBSR}(a), the corresponding curves of the described one-parameter and two-parameter OPE fits. In Fig.~\ref{FigBSRECHvsOPE} we present, for comparison, these OPE-curves (the central cases) together with the pure ECH curve of Fig.~\ref{FigBSR}(a).
\endgroup

\begingroup\color{black}
\section{Summary}
\label{sec:summ}

In this work we constructed an extension $\A_{\rm ECH}(Q^2) = \A(\kappa_{\rm ECH} Q^2, c_2^{\rm ECH}, c_3^{\rm ECH})$ of the pQCD coupling $a_{\rm ECH}(Q^2) = a(\kappa_{\rm ECH} Q^2, c_2^{\rm ECH}, c_3^{\rm ECH})$ in the ($N_f=3$ four-loop) ECH scheme for BSR canonical part $d(Q^2)$. This extension simultaneously fulfills several conditions:
\begin{enumerate}
\item  
In contrast to the pQCD coupling, $\A_{\rm ECH}(Q^2)$ is a holomorphic coupling, i.e., it has no Landau singularities, i.e., no singularities in the (generalised) Euclidean part of the complex $Q^2$-plane, $Q^2 \in \mathbb{C} \backslash (-\infty, 0]$. This coupling thus reflects the known holomorphic properties of the QCD spacelike observables (such as BSR, Adler function, etc.).
\item
This coupling practically coincides with the corresponding pQCD coupling $a_{\rm ECH}(Q^2)$ at high $|Q^2| > 1 \ {\rm GeV}^2$, namely $\A_{\rm ECH}(Q^2) - a_{\rm ECH}(Q^2) \sim (\Lambda^2_{\rm QCD}/Q^2)^5$.
\item  
  At the limit $Q^2 \to 0$, the values of the coupling  $\A_{\rm ECH}(Q^2)$ and its first two derivatives coincide with the predictions of the Holographic Light-Front QCD (HLFQCD) effective coupling \cite{LFH1} $\A(Q^2)_{\rm HLF} = \exp(-Q^2/\tkap^2)$ (where $\tkap \approx 0.523$ GeV).
\end{enumerate}
The holomorphic coupling is constructed on the premise that its discontinuity (spectral) function $\rho_{\rm A}(\sigma) = {\rm Im} \; \A(-\sigma - i \epsilon; c_2^{\rm ECH}, c_3^{\rm ECH})$ at large enough (timelike) squared energies $\sigma \geq M_0^2$ ($=s_0 \LL^2 \approx 3.2-5.6 \ {\rm GeV}^2$) coincides with the corresponding pQCD spectral function $\rho_a(\sigma) = {\rm Im} \; a(-\sigma - i \epsilon; c_2^{\rm ECH}, c_3^{\rm ECH})$, and at lower (timelike) squared energies ($ 0 < \sigma < M_0^2$) its otherwise unknown behaviour is parametrised by a combination of Dirac delta functions and their derivatives.

Numerical comparison of the obtained ECH coupling with the BSR experimental data shows a very good agreement in the entire considered ($N_f=3$) spacelike $Q^2$-interval $0 < Q^2 < 4.74 \ {\rm GeV}^2$.  

\textcolor{black}{For comparison, we also performed OPE fit analysis analogous to the analysis in our previous work \cite{pNPB}, but this time using in the leading-twist ($D=0$) part the considered expressions of our pure ECH approach Eqs.~(\ref{BSRECH}), (\ref{dECH}). This analysis, however, gives results that break down at low $Q^2$, due to the usual singular behaviour of the OPE terms there.}
  
Our mathematica programs that evaluate $\A(Q^2)$, for the three choices of the spectral function threshold scales, $M_1= 0.279^{+0.116}_{-0.065}$ GeV, are available on www \cite{www}.

\begin{acknowledgments}
This work was supported in part by Proyecto UTA Mayor No.~6752-24, y FONDECYT (Chile) Grants No.~1240329 (C.A.) and No.~1220095 (G.C.). 
\end{acknowledgments}
\endgroup

\appendix

\begingroup\color{black}
\section{Renormalisation scheme dependence}
\label{app:RSch}

The renormalisation scheme dependence (briefly: scheme) of the coupling $a \equiv a(\kappa Q^2; c_2, \ldots)$ and of the coefficients $d_n \equiv d_n(\kappa; c_2, \ldots)$ is the dependence on the RScl parameter $\kappa$ and on the beta-coefficients $c_j \equiv \beta_j/\beta_0$ ($j=2,3, \ldots$) appearing in the RGE Eq.~(\ref{RGE}). The scheme dependence of the mentioned running coupling $a$ is governed by the RGE (\ref{RGE}) and the following relations (cf.~App.~A of \cite{Stevenson}, and App.~A of \cite{GCRK63}):
\bes
\label{aRSch}
\bea
\frac{\partial a}{\partial c_2} & = & a^3 + \frac{c_2}{3} a^5 + {\cal O}(a^6),
\label{ac2} \\
\frac{\partial a}{\partial c_3} & = & \frac{1}{2}a^4 - \frac{c_1}{6} a^5 + {\cal O}(a^6),
\label{ac3} \\
\frac{\partial a}{\partial c_4} & = & \frac{1}{3} a^5 + {\cal O}(a^6).
\label{ac4} \eea \ees
When the RGE (\ref{RGE}) and these relations are used in the power expansion Eq.~(\ref{dptkap}) of the canonical ($D=0$) BSR $d(Q^2)$, and we take into account that $d(Q^2)$ is scheme-independent (i.e., independent of $\kappa, c_2, c_3, \ldots$), we obtain the explicit scheme-dependence of the expansion coefficients $d_n \equiv d_n(\kappa; c_2, c_3, \ldots)$ in terms of the coefficients in the reference scheme  ${\bar d}_n \equiv d_n(1; {\bar c}_2,\ldots,{\bar c}_n)$ (i.e., the canonical $\kappa=1$ $\MSbar$ scheme)
\bes
\label{dnRSch}
\bea
d_1 & = & {\bar d}_1 + \beta_0 \ln \kappa,
\label{d1RSch}
\\
d_2 &=& {\bar d}_2 + (\beta_0 \ln \kappa) (2 {\bar d}_1 + c_1) + (\beta_0 \ln \kappa)^2 - (c_2 - {\bar c}_2),
\label{d2RSch}
\\
d_3 & = & \left[ {\bar d}_3 - 2 (c_2 - {\bar c}_2) {\bar d}_1 - \frac{1}{2} (c_3 - {\bar c}_3) \right] + (\beta_0 \ln \kappa) \left[ 3 {\bar d}_2 + 2 c_1 {\bar d}_1 -  2 (c_2 - {\bar c}_2) + {\bar c}_2 \right]
\nonumber\\ &&
+ (\beta_0 \ln \kappa)^2 \left( 3 {\bar d}_1 + \frac{5}{2} c_1 \right) +  (\beta_0 \ln \kappa)^3.
\label{d3RSch}
\eea \ees

The (perturbative) four-loop ECH scheme is then characterised by $\kappa_{\rm ECH}$, $c_2^{\rm ECH}$ and $c_3^{\rm ECH}$ such that $d_j^{\rm ECH}=0$ for $j=1,2,3$. The above relations then give immediately the values of these ECH parameters in terms of the coefficients ${\bar d}_j$ in the reference scheme
\bes
\label{ECHval}
\bea
\beta_0 \ln \kappa_{\rm ECH} &=& -{\bar d}_1 \quad \left( \kappa_{\rm ECH} = \exp\left(-{\bar d}_1 /\beta_0 \right) \right),
\label{kapECH} \\
c_2^{\rm ECH} & = & {\bar c}_2 + \left[ {\bar d}_2 - c_1 {\bar d}_1 - ({\bar d}_1)^2 \right],
\label{c2ECH}
\\
c_3^{\rm ECH} & = & {\bar c}_3 + 2 \left[ {\bar d}_3 - 3 {\bar d}_2 {\bar d}_1 + \frac{1}{2} c_1 ({\bar d}_1)^2 + 2 ({\bar d}_1)^3 - {\bar c}_2 {\bar d}_1 \right].
\label{c3ECH}
\eea \ees

\section{P44 renormalisation schemes}
\label{app:P44}

In the RGE (\ref{RGE}), the scheme parameters are $c_j$ ($j=2,3,\ldots$). The considered ECH scheme is based on the available knowledge of all the exactly known BSR perturbation expansion coefficients $d_n$ ($n=1,2,3$). Therefore, as seen, only two $c_j$ coefficients can be fixed in this scheme, namely $c_2$ and $c_3$ (their numerical values are given in the text). Therefore, in principle, the RGE-evolution of the (pQCD) ECH coupling $a(\mu^2)$ could be regarded as governed by the following four-loop truncated beta-function:
\be
\frac{d}{d \ln \mu^2} a(\mu^2) = - \beta_0 a(\mu^2)^2 \left[ 1 + c_1 a(\mu^2) + c_2 a(\mu^2)^2 + c_3 a(\mu^2)^3 \right]
\label{RGEc2c3} \ee
This differential equation has no explicit solution in terms of known functions, and can thus be solved (in the considered $N_f=3$ regime) only numerically.\footnote{We consider that we know an initial condition there, e.g., the value of $a({\overline m}_c^2)$.} To evaluate the holomorphic coupling $\A(Q^2)$ whose underlying pQCD couopling is this $a(Q^2)$, we need to apply the dispersion relation (\ref{Adisp}). In this dispersive integral enters as integrand the spectral (or: discontinuity) function $r_{a}(s) = \rho_a(\sigma) = {\rm Im} \; a(-\sigma - i \epsilon)$ (where: $s \equiv \sigma/\LL^2$) for an entire $s$-interval $s_0 \leq s < \infty$ ($s_0>0$), cf.~Eqs.~(\ref{rA}) and (\ref{Aexpr}). This means, in practice, that we would have to evaluate numerically this spectral function $r_{a}(s)$ in an (almost) infinite $s$-interval with a good precision, which is practically impossible.\footnote{For the numerical evaluation of $r_{a}(s) = {\rm Im} \; a(-s \LL^2 - i \epsilon)$, we need to evaluate numerically $a(Q^2)$ close to the negative $Q^2$-axis (cut).  We note that the numerical integration of the RGE starts failing when we approach the singularities (cuts) of $a(Q^2)$.}
  Therefore, we will restrict ourselves to a class of schemes, namely the so called P44-class, which allows an explicit (and thus convenient) solution of the RGE (\ref{RGE}) for the running coupling $a(Q^2)$, and whose beta-function agrees up to four-loop with the beta-function Eq.~(\ref{RGEc2c3}). The beta-function in this P44-class has only two adjustable scheme parameters, namely the two leading scheme parameters $c_2$ and $c_3$ (while $c_j$'s for $j \geq 4$ are then specific functions of $c_2$ and $c_3$). Such a beta-function $\beta(a)$ has a diagonal Pad\'e form [4/4](a) ('P44'), i.e., it is a ratio of two  polynomials of degree 4 in $a(Q^2)$
\be
\frac{d a(Q^2)}{d \ln Q^2}  =
\beta(a(Q^2))
\equiv
- \beta_0 a(Q^2)^2 \frac{ \left[ 1 + \alpha_0 c_1 a(Q^2) + \alpha_1 c_1^2 a(Q^2)^2 \right]}{\left[ 1 - \alpha_1 c_1^2 a(Q^2)^2 \right] \left[ 1 + (\alpha_0-1) c_1 a(Q^2) + \alpha_1 c_1^2 a(Q^2)^2 \right]} \ ,
\label{beta} \ee
where $c_j \equiv \beta_j/\beta_0$ and
\be
\alpha_0  =  1 + \sqrt{c_3/c_1^3}, \quad 
\alpha_1 = c_2/c_1^2 +   \sqrt{c_3/c_1^3} .
\label{a0a1} \ee
When we expand this $\beta$-function in powers of $a(Q^2)$, the terms up to the (four-loop) term with $c_3$ of the expansion (\ref{RGE}) are reproduced, while the terms with $c_j$ ($j \geq 4$) have the coefficients $c_j$ as specific functions of $c_2$ and $c_3$. The RGE (\ref{beta}) has explicit solution in terms of the Lambert functions $W_{\mp 1}(z)$, as shown in \cite{GCIK}
\be
a(Q^2) = \frac{2}{c_1}
\left[ - \sqrt{\omega_2} - 1 - W_{\mp 1}(z) + 
\sqrt{(\sqrt{\omega_2} + 1 + W_{\mp 1}(z))^2 
- 4(\omega_1 + \sqrt{\omega_2})} \right]^{-1},
\label{aP44} \ee
where $\omega_1= c_2/c_1^2$, $\omega_2=c_3/c_1^3$,  $Q^2 = |Q^2| \exp(i \phi)$, and $W_{\mp 1}(z)$ are two branches of the Lambert function. When $0 \leq \phi < \pi$, $W_{-1}(z)$ is used; when $-\pi \leq \phi < 0$, $W_{+1}(z)$ is used. The argument $z = z(Q^2)$ appearing in $W_{\pm 1}(z)$ is
\be
z \equiv z(Q^2) =
-\frac{1}{c_1 e}
\left( \frac{\LL^2}{Q^2} \right)^{\beta_0/c_1} \ .
\label{zexpr} \ee
We call the scale $\LL$ appearing here the Lambert scale; it turns out that $\LL^2 \sim \Lambda^2_{\rm QCD}$ ($\sim 0.01$-$0.1 \ {\rm GeV}^2$). The scale convention in all these schemes is the same as in $\MSbar$, only the chosen scheme parameters ($c_2$, $c_3$) are now in general different from those in $\MSbar$. The scale $\LL$ is related to the strength of the coupling. We will call the described class of schemes as P44-schemes. They are all for the $N_f=3$ case, i.e., QCD with three massless quarks (and the other quarks are considered decoupled).

The strength of the coupling Eq.~(\ref{aP44}), or equivalently the scale $\LL$, is determined by the value of $\alpha_s^{\MSbar}(M_Z^2)$. This is obtained in the following way. We RGE-evolve $a(Q^2) \equiv \alpha_s(Q^2)/\pi$ from $Q^2=M_Z^2$ (where $N_f=5$) with the five-loop $\MSbar$ RGE \cite{5lMSbarbeta} downwards, and take the corresponding four-loop quark threshold relations \cite{4lquarkthr1,4lquarkthr2} at $Q^2 = k {\overline m}_q$ (we take $k=2$; ${\overline m}_b=4.2$ GeV; ${\overline m}_c=1.27$ GeV). Then at a scale $Q^2=Q_0^2$ and $N_f=3$ [we took $Q_0^2 = (2 {\overline m}_c)^2$] we change the scheme from the five-loop $\MSbar$ to the mentioned P44-scheme with chosen $c_2$ and $c_3$ values, via the relation (cf.~App.~A of \cite{Stevenson} and App.~A of \cite{GCRK63})
\bea 
\lefteqn{
  \frac{1}{a} + c_1 \ln \left( \frac{c_1 a}{1 + c_1 a} \right) + \int_0^a
  dx \left[ \frac{\beta(x) + \beta_0 x^2 (1 + c_1 x)}{x^2 (1 + c_1 x) \beta(x)}
    \right]
}
\nonumber \\
& & =
\frac{1}{{\bar a}} + c_1 \ln \left( \frac{c_1 {\bar a}}{1 + c_1 {\bar a}} \right) + \int_0^{\bar a}
  dx \left[ \frac{{\overline \beta}(x) + \beta_0 x^2 (1 + c_1 x)}{x^2 (1 + c_1 x) {\overline \beta}(x)}
    \right],
  \label{abara} \eea
  where $a= a(Q_0^2)$ is the coupling in the chosen (P44) scheme, ${\bar a}={\bar a}(Q_0^2)$ is the coupling in the five-loop $\MSbar$ scheme obtained by the aforementioned RGE-evolution, and ${\overline \beta}(x)$ is the five-loop $\MSbar$ beta-function (polynomial), all with $N_f=3$. The above relation (\ref{abara}) is solved numerically to obtain the value of $a = a(Q_0^2)$ in the P44 scheme with chosen $c_2$ and $c_3$. From here, using Eq.~(\ref{aP44}) (with $Q^2=Q_0^2$) and the relation (\ref{zexpr}), we obtain the value of $\LL$, namely $\Lambda_{\rm L}= 0.078107$ GeV in the ($N_f=3$) ECH scheme when $\alpha_s^{\MSbar}(M_Z^2)=0.1179$. Therefore, the value of the scaling parameter $\LL$ is just the reflection of the chosen (input) value $\alpha_s^{\MSbar}(M_Z^2)$ (and of the chosen scheme parameters $c_2$ and $c_3$).\footnote{Specifically, when $\alpha_s^{\MSbar}(M_Z^2)=0.1179$, we obtain in the regime $N_f=3$ in the (four-loop) ECH scheme $\LL= 0.078107$ GeV.} Then we can obtain $a(Q^2)$ at any other $Q^2$ (and keeping $N_f=3$) by the formula (\ref{aP44}).

The required spectral function $r_{a}(s)= {\rm Im} \; a(-s \LL^2 - i \epsilon)$ can then be evaluated directly by using the formula (\ref{aP44}).

\section{The high- and low-momentum conditions for the coupling $\A(Q^2)$}
\label{app:Acond}

The pQCD coupling $a(Q^2)$ has not only the (physically) expected singularities along the negative (i.e., timelike) $Q^2$-axis, but also artificial (Landau) singularities along the positive (i.e., spacelike) $Q^2$-axis, $0 \leq Q^2  \leq \sigma_{\rm c}$. This means that its spectral function $r_{a}(s)= {\rm Im} \; a(-s \LL^2 - i \epsilon)$ has nonzero values not just for $s>0$, but also for negative values $-s_{\rm c} \leq s \leq 0$  (where $s_{\rm c} = \sigma_{\rm c}/\LL^2$). This is illustrated in Fig.~\ref{figtr1}.
\begin{figure}[htb] 
\begin{minipage}[b]{.49\linewidth}
\includegraphics[width=85mm,height=50mm]{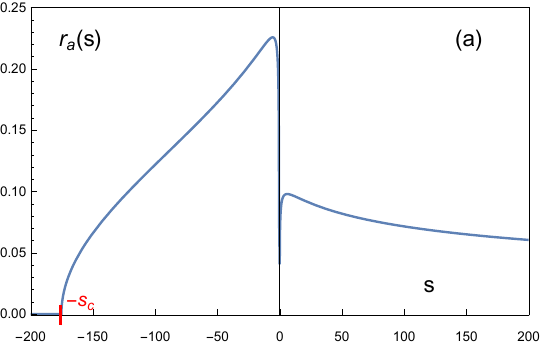}
\end{minipage}
\begin{minipage}[b]{.49\linewidth}
  \includegraphics[width=85mm,height=50mm]{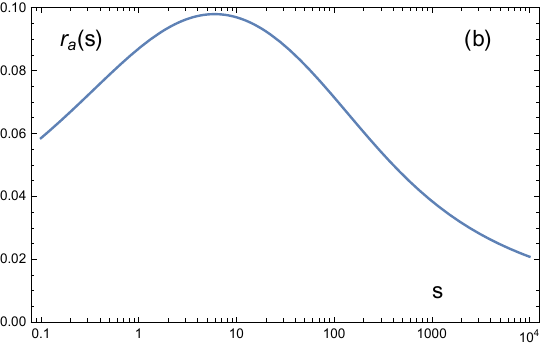}
\end{minipage} 
\caption{\footnotesize The spectral function $r_{a}(s)= {\rm Im} \; a(-s \LL^2 - i \epsilon)$ of the pQCD coupling in the ECH P44 scheme in the $N_f=3$ regime; $\LL=0.078107$ GeV corresponding to $\alpha_s^{\MSbar}(M_Z^2)=0.1179$. (a) $r_{a}(s)$ in low-$s$ regime, including the Landau cut region ($-s_{\rm c} \leq s \leq 0$) where $s_{\rm c} \approx 176$, i.e., $\sigma_{\rm c}=s_{\rm c} \LL^2 \approx 1.07 \ {\rm GeV}^2$; (b) $r_{a}(s)$ for positive $s$, including large-$s$. Fig.~(b) is on the logarithmic $s$-scale.}
\label{figtr1}
\end{figure}

The high-momentum ($|Q^2| > 1 \ {\rm GeV}^2$) conditions, Eqs.~(\ref{diffAa}), are in fact four conditions, because the difference is $\A(Q^2) - a(Q^2) \sim 1/Q^2$ (at high $|Q^2|$) by default. We enforce these conditions for the dispersive expression of $\A(Q^2)$ Eq.~(\ref{Aexpr}) by expanding the expression (\ref{Aexpr}) in inverse powers of $u$ ($\equiv Q^2/\LL^2$) up to $1/u^5$. Application of the conditions (\ref{diffAa}) to this expansion then gives the four (high-momentum) conditions
\bes
\label{highm}
\bea
\sum_{j=1}^2 f_j^{(0)} &=& P^{(0)},
\label{P0} \\
\sum_{j=1}^2 f_j^{(0)} s_j - f_1^{(1)} & = & P^{(1)},
\label{P1} \\
\sum_{j=1}^2 f_j^{(0)} s_j^2 - 2 s_1 f_1^{(1)} + 2 f_1^{(2)} & = & P^{(2)},
\label{P2} \\
\sum_{j=1}^2 f_j^{(0)} s_j^3 - 3 s_1^2 f_1^{(1)} + 6 s_1f_1^{(2)} - 6 f_1^{(3)} & = & P^{(3)},
\label{P3}
\eea \ees
where the right-hand sides $P^{(j)}$ are the contributions from the (spacelike) Landau cut ($-s_{\rm c} \leq s \leq 0$) and the low-momentum (timelike) regime up to the ``pQCD-onset'' scale $s_0$  ($0 < s \leq s_0$)
\be
P^{(j)} \equiv \frac{1}{\pi} \int_{-s_{\rm c}}^{s_0} d s \; s^j r_{a}(s).
\label{Pjs} \ee

On the other hand, the low-momentum ($Q^2 \to 0$) conditions (\ref{AQ0}) are implemented in the dispersive expression (\ref{Aexpr}) directly, i.e., by simply applying this expression, and the first and second derivatives of it, at $u=0$
\bes
\label{lowm}
\bea
\sum_{k=0}^{3} \frac{f_1^{(k)} k!}{s_1^{k+1}} + \frac{f_2^{(0)}}{s_2} & = & 1 - B^{(1)},
\label{Pm1} \\
\sum_{k=0}^{3} \frac{f_1^{(k)} (k+1)!}{s_1^{k+2}} + \frac{f_2^{(0)}}{s_2^2} & = & +\frac{\LL^2}{4 \tkap^2 \kappa_{\rm ECH}} - B^{(2)},
\label{Pm2} \\
\frac{1}{2} \sum_{k=0}^{3} \frac{f_1^{(k)} (k+2)!}{s_1^{k+3}} + \frac{f_2^{(0)}}{s_2^3} & = & + \frac{1}{2} \left( \frac{\LL^2}{4 \tkap^2 \kappa_{\rm ECH}} \right)^2 - B^{(3)},
\label{Pm3}
\eea \ees
Here, $B^{(j)}$ are the expressions
\be
B^{(j)} \equiv \frac{1}{\pi} \int_{s_0}^{\infty} \frac{d s}{s^j} r_{a}(s).
\label{Bjs} \ee

We recall that $\tkap \approx 0.523$ GeV \cite{PLB759} as mentioned in the text, and $\kappa_{\rm ECH}=0.203398$ is the (four-loop) ECH rescaling parameter, cf.~Eq.~(\ref{RSchECH}).

In Table \ref{TabParam} we present the obtained values of the parameters of the model, for the three choices of the $M_1$-threshold scale $M_1=(0.279^{+0.116}_{-0.065})$ GeV.
\begin{table}
\caption{The values of the parameters of the considered AQCD coupling, cf.~Eqs.~(\ref{rA})-(\ref{Af}) and the conditions (\ref{highm})-(\ref{Bjs}). We note that with the choice $\alpha_s^{\MSbar}(M_Z^2)=0.1179$, the Landau scale is $\Lambda_{\rm L} = 0.078107$ GeV. The values are presented for the three considered values of the threshold scale $M_1= \sqrt{s_1} \Lambda_{\rm L}$: $M_1=(0.279^{+0.116}_{-0.065})$ GeV.}
\label{TabParam}
\begin{ruledtabular}
\centering
\begin{tabular}{r|llllllll}
 $M_1$ (GeV) & $s_1$ & $s_2$ & $f_1^{(0)}$ & $f_1^{(1)}$ & $f_1^{(2)}$ & $f_1^{(3)}$ & $f_2^{(0)}$ & $s_0$ 
\\
\hline
0.2792 & 12.7777 & 606.836 & 13.9358 & 139.574 & -1761.09 &  2903.86 & 7.96876 & 847.7 \\
\hline
0.3946 & 25.5224 & 335.593 & 9.24753 & 1123.48 & -13823.9 & 33139.9 & 8.34028 & 532.
\\
0.2136 & 7.48085 & 657.268 & 14.4678 & -63.0469 & -21.8665 & 109.279 & 8.26638 & 913.
\\
\hline
\end{tabular}
\end{ruledtabular}
\end{table}

\section{The charm mass nondecoupling contribution $\delta d(Q^2)_{m_c}$}
\label{app:ddmc}

These effects were evaluated in \cite{Blumetal}. We neglect the (heavy) $b$-quark contributons (i.e., we consider $m_b \to \infty$). Then the considered effects, at next-to-leading order (NLO, $\sim a^2$), are expressed with  the function $C_{\rm pBJ}^{\rm mass.,(2)}(\xi_c)$. Here,  $\xi_c \equiv Q^2/m_c^2$ where $m_c \approx 1.67$ GeV is the pole mass. The function $C_{\rm pBJ}^{\rm mass.,(2)}$ appears originally in the coefficient at $a^2$ when the perturbation expansion (\ref{dptkap}) with $\kappa=1$, which is for $N_f=3$ coupling, is expressed in terms of the $N_f=4$ coupling $a(Q^2)_{N_f=4}$
\bea
d(Q^2)_{\rm pt} &=& a(Q^2)_{N_f=4} + a(Q^2)^2_{N_f=4}
\left\{ \frac{55}{12} - \frac{1}{3} \left[ N_f -1 + C_{\rm pBJ}^{\rm mass.,(2)}(\xi_c) \right] \right\} + {\cal O}(a^3).
\label{dmc1}
\eea
Here, $N_f=4$ and the expression for $C_{\rm pBJ}^{\rm mass.,(2)}(\xi_c)$ is  
\bea
C_{\rm pBJ}^{\rm mass.,(2)}(\xi) & = & \frac{1}{2520} {\bigg \{} \frac{1}{\xi} ( 6 \xi^2 + 2735 \xi + 11724 ) - \frac{\sqrt{\xi+4}}{\xi^{3/2}} ( 3 \xi^3 + 106 \xi^2 + 1054 \xi + 4812) \ln \left[ \frac{ \sqrt{\xi + 4} + \sqrt{\xi} }{\sqrt{\xi + 4} - \sqrt{\xi} } \right]
\nonumber\\
&&
- 2100 \frac{1}{\xi^2} \ln^2 \left[ \frac{ \sqrt{\xi + 4} + \sqrt{\xi} }{\sqrt{\xi + 4} - \sqrt{\xi} } \right] + (3 \xi^2 + 112 \xi + 1260) \ln \xi {\bigg \}}.
\label{CBj1}
\eea
We note that when $Q^2 \gg m_c^2$ ($\xi \gg 1$), this function approaches unity quite slowly
\be
C_{\rm pBJ}^{\rm mass.,(2)}(\xi)  = 1 - \frac{8}{3} \frac{\ln \xi}{\xi} + \frac{34}{9\xi }  + {\cal O}\left(  \frac{\ln^2 \xi}{\xi^2} \right),
\label{CBj2}
\ee
and in this limit we obtain the massless $N_f=4$ QCD expression for $d_1$
\be
d_1(N_f)= \frac{55}{12} - \frac{1}{3} N_f
\label{e2NS}
\ee
with $N_f=4$.

We now reexpress in Eq.~(\ref{dmc1}) the coupling $a(Q^2)_{N_f=4}$ with our used coupling $a(Q^2)$ ($\equiv a(Q^2)_{N_f=3}$)
(cf.~e.g.,~\cite{CKS})
\be
a(Q^2)_{N_f=4} = a(Q^2) + \frac{1}{6} \ln \left( \frac{Q^2}{m_c^2} \right) a(Q^2)^2 + {\cal O}(a^3),
\label{aNf4vsa} \ee
and this leads to the following pQCD expression for the correction $\delta d(Q^2)_{m_c}$:
\be
\delta d(Q^2)_{m_c} =
\frac{1}{6} \left[ \ln \left( \frac{Q^2}{m_c^2} \right) -
  2 C^{\rm mass.,(2)}_{\rm pBJ}\left( \frac{Q^2}{m_c^2} \right) \right] a(Q^2)^2 +
{\cal O}(a^3).
\label{deldp} \ee
This should be interpreted as the correction due to nondecoupling of the charm quark (i.e., due to $m_c \not= \infty$). We note that $d_1$ in our perturbation expansion (\ref{dptkap}) is for $d(N_f)$ of Eq.~(\ref{e2NS}) for $N_f=3$, i.e., in the massless $N_f=3$ QCD.

In AQCD variants, where $a(Q^2) \mapsto \A(Q^2)$, we do not have $a(Q^2)^2 \mapsto \A(Q^2)^2$, because analytisation can be applied consistently only to expressions that are linear in $a(Q^2)$. However, the logarithmic derivative
\be
\ta_2(Q^2) \equiv \frac{(-1)}{\beta_0} \frac{d}{d \ln Q^2} a(Q^2)
\label{ta2}
\ee
is linear in $a(Q^2)$, and perturbatively $\ta_2 = a^2 + {\cal O}(a^3)$. Therefore, in our AQCD approach we replace the power $a(Q^2)^2$ in Eq.~(\ref{deldp}) simply by the analytised version of $\ta_2(Q^2)$, i.e., by
\be
\ta_2(Q^2) \mapsto \tA_2(Q^2) \equiv \frac{(-1)}{\beta_0} \frac{d}{d \ln Q^2} \A(Q^2),
\label{tA2}
\ee
leading to our final result for $\delta d(Q^2)_{m_c}$
\be
\delta d(Q^2)_{m_c} = d_1(Q^2)_{m_c} \tA_2(Q^2) =
\frac{1}{6} \left[ \ln \left( \frac{Q^2}{m_c^2} \right) -
  2 C^{\rm mass.,(2)}_{\rm pBJ}\left( \frac{Q^2}{m_c^2} \right) \right] \tA_2(Q^2).
\label{deldA} \ee
Since the terms $\sim a^3$ ($\sim \ta_3 \sim \tA_3$) in this expression are not known, it does not matter in which scheme we evaluate $\tA_2(Q^2)$, so we evaluated it in the considered (P44) ECH scheme: $\tA_2(Q^2; c_2^{\rm ECH}, c_3^{\rm ECH})$.

In general, the evaluation of (truncated) perturbation series of QCD observables in powers of $a(Q^2)$ is performed in AQCD variants in the following way. First we reorganise the expansion in powers $a^n$ into the series in terms of the logarithmic derivatives $\ta_n(Q^2) \propto (d/d \ln Q^2)^{n-1} a(Q^2)$, and then replace these derivatives with their AQCD counterparts, $\tA_n(Q^2) \propto (d/d \ln Q^2)^{n-1} \A(Q^2)$. This construction was introduced in \cite{CV}, where also the expressions $\A_n(Q^2)$ [analogs of powers $a(Q^2)^n$] were constructed.  We stress that this approach of analytisation is unambiguous, once we have a given (AQCD) coupling $\A(Q^2)$. The extension of $\tA_{\nu}(Q^2)$ and $\A_{\nu}(Q^2)$ to noninteger $\nu$ ($-1 < \nu$) was performed in \cite{GCAK}.\footnote{
For the specific case of the Minimal Analytic (MA) QCD \cite{Shirkov,SMS,ShirRev}, the extended logarithmic derivatives $\tA_{\nu}^{\rm (MA)}(Q^2)$ were constructed as explicit functions at one-loop order in \cite{FAPT} and at any loop order in \cite{Kotikov}.}

In Figs.~\ref{Figddmc} we present the $Q^2$-dependence of the coefficient $d_1(Q^2)_{m_c}$ that appears in Eq.~(\ref{deldA}) and the full expression of  Eq.~(\ref{deldA}). It is clear that the coefficient goes to zero fast when $Q^2 \to 0$, and $\delta d(Q^2)_{m_c}$ is very small and converges at small $Q^2$ to zero even faster than the coefficient when $Q^2 \to 0$ (and $Q^2 < 0.4 \ {\rm GeV}^2$).
\begin{figure}[htb] 
\begin{minipage}[b]{.49\linewidth}
\includegraphics[width=85mm,height=50mm]{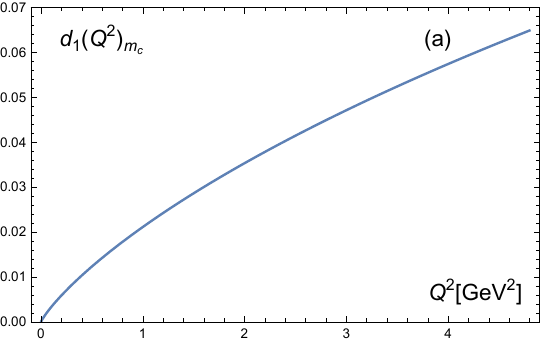}
\end{minipage}
\begin{minipage}[b]{.49\linewidth}
  \includegraphics[width=85mm,height=50mm]{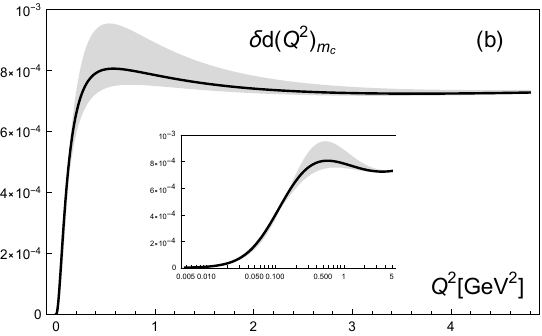}
\end{minipage} 
\caption{\footnotesize (a) The coefficient $d_1(Q^2)_{m_c}$ that appears in Eq.~(\ref{deldA}), as a function of $Q^2$; (b) the contribution $\delta d(Q^2)_{m_c}$ Eq.~(\ref{deldA}), for the three threshold scales used in $\A(Q^2)$: $M_1 = (0.279^{+0.116}_{-0.065})$ GeV. The solid line is for the central value $M_1=0.279$ GeV. The upper and the lower border of the grey area are for $M_1=0.395$ GeV and $M_1=0.214$ GeV, respectively. The corresponding curve with $Q^2$ scaled logarithmically is included.}
\label{Figddmc}
\end{figure}
\endgroup

\end{document}